\documentclass[aps,prb,amsmath,amssymb,twocolumn,superscriptaddress]{revtex4-2}
\pdfoutput=1 
\usepackage[usenames]{xcolor}
\usepackage[pdftex]{graphicx}
\usepackage{amsmath, amstext, amssymb, amsfonts, amsxtra}
\usepackage{xspace}
\usepackage{blindtext}
\usepackage{braket}
\usepackage{dsfont}
\usepackage{placeins}
\usepackage{here}
\usepackage[colorlinks=true,linkcolor=black,citecolor=blue,urlcolor=black]{hyperref}
\usepackage{siunitx}

\begin{document}
\title{Heat transport in a Coulomb ion crystal with a topological defect}
\author{L. Timm}
\email[]{lars.timm@itp.uni-hannover.de}
\affiliation{Institut f\"ur Theoretische Physik, Leibniz Universit\"at Hannover, Appelstr. 2, 30167 Hannover, Germany}
\author{H. Weimer}
\affiliation{Institut f\"ur Theoretische Physik, Technische Universit\"at Berlin, Hardenbergstra{\ss}e 36 EW 7-1, 10623 Berlin, Germany}
\author{L. Santos}
\affiliation{Institut f\"ur Theoretische Physik, Leibniz Universit\"at Hannover, Appelstr. 2, 30167 Hannover, Germany}
\author{T. E. Mehlst\"aubler}
\affiliation{Physikalisch-Technische Bundesanstalt, Bundesallee 100, 38116 Braunschweig, Germany}
\affiliation{Institut f\"ur Quantenoptik, Leibniz Universit\"at Hannover, Welfengarten 1, 30167 Hannover,
Germany}

\date{\today}

\begin{abstract}

The thermodynamics of low-dimensional systems departs significantly from phenomenologically deducted macroscopic laws. 
Particular examples, not yet fully understood, are provided by the breakdown of Fourier's law and the ballistic transport of heat.  
Low-dimensional trapped ion systems provide an experimentally accessible and well-controlled platform for the study of these problems. 
In our work, we study the transport of thermal energy in low-dimensional trapped ion crystals, focusing in particular on the influence of the Aubry-like transition that occurs when a topological defect is present in the crystal.
We show that the transition significantly hinders efficient heat transport, being responsible for the rise of a marked 
temperature gradient in the non-equilibrium steady state. 
Further analysis reveals the importance of the motional eigenfrequencies of the crystal.
\end{abstract}

\maketitle



\section{Introduction}

The study of how heat is transported through a system in different phases is an active field of research since the days of Newton and Fourier~\cite{fourier2009,newton1701,schroedinger1914,born1921,peierls1929}. 
Surprisingly, well-established phenomenological findings seem to be invalid in low-dimensional harmonic systems~\cite{dhar2008}.
Phenomena contradictory to the laws governing the heat transport on macroscopic scales have been observed in different lattice models~\cite{dhar2008, lepri2003}, sparking the interest to understand the role of different microscopic properties.
The interplay of linear and nonlinear dynamics, as in the well-known Fermi-Pasta-Ulam model~\cite{fermi1955}, or the importance of integrability and disorder in the system, are some examples to be named here~\cite{hu1998,lepri1997,hu2005,savin2003,roy2008,wang2020}. 

While transport in these theoretical models, also in the quantum realm~\cite{bermudez2016,pekola2021}, has attracted considerable attention, the experimental investigation of low-dimensional systems proved to be difficult due to the lack of a well suited platform with sufficient control and readout techniques. In this context, trapped ions offer a particularly interesting platform, with excellent access to the particles, as well as a rich variety of laser manipulation and readout techniques~\cite{drewsen2015}. 
In addition, the possibility to vary the confinement in the different trap dimensions allows for the tuning of the geometry and the dimensionality of the crstals, nonlinear effects are introduced due to the Coulomb interaction.



\begin{figure}
    \centering
    \includegraphics[width=0.45\textwidth]{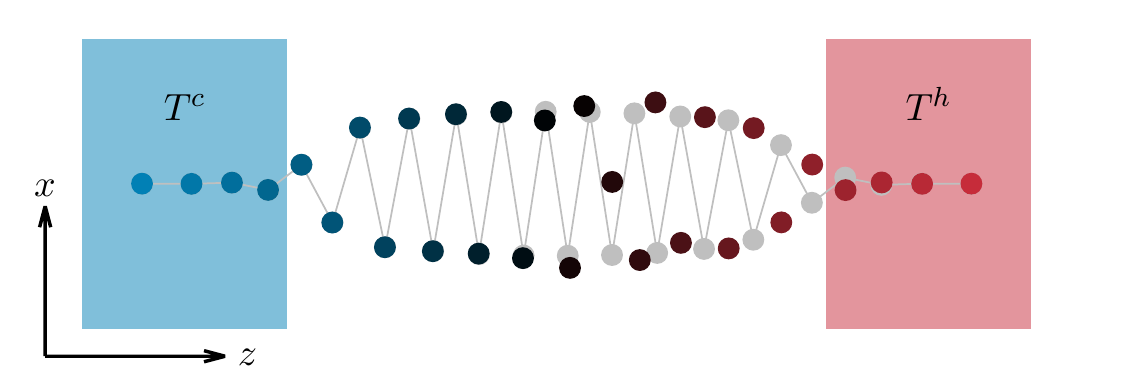}
    \caption{Schematic depiction of the system considered. The outer four ions of a zigzag crystal with a kink defect are coupled to Langevin heat baths with different temperatures. The ion crystal develops a temperature profile in the steady state. The grey points depict the zigzag crystal, the coloured points are for a crystal with a localized (odd) defect in the central region.}
    \label{fig:scheme}
\end{figure}


Along this direction, there have been already several theoretical works studying the transport of energy in ion crystals in different configurations and limits~\cite{lin2011,ramm2014,ruiz2014,freitas2016,mao2022,timm2020}, while first experiments demonstrated the controlled insertion of motional excitations and the readout of their dynamics through the crystal~\cite{abdelrahman2017,tamura2020}. In addition to regular lattice configurations, such as linear chains or triangular lattices, experiments have realized stable lattice defects in ion crystals~\cite{pyka2013,mielenz2013,landa2013}.
Their discovery has triggered a series of investigations, including the experimental confirmation of the Kibble-Zurek scaling for their creation probability, and their ability to emulate paradigmatic models of nanofriction~\cite{pyka2013,delcampo2010,ulm2013,partner2015,kiethe2017,kiethe2018}.
Especially the detection of a sliding-to-pinned transition, celebrated in the tribology context as the Aubry transition \cite{aubry1983}, has opened questions regarding their effect on the dynamics of local excitations~\cite{toda1979,hu2005,tsironis1999}. 
Simultaneously, a complementary approach employing linear ion strings in the periodic potential of a standing wave laser field has demonstrated similar physics, showing the existence of a frictionless phase ~\cite{garcia-mata2007,bylinskii2015,bylinskii2016,gangloff2020,bonetti2021}.

Previous work showed that energy transport in ion crystals in the presence of a topological defect is nontrivial~\cite{timm2020}. 
In this paper, we expand this study in the context of thermal conductivity and investigate heat flux through the crystal and the defect.
In particular, we are interested in how the robust energy localization observed in the pinned phase translates into the temperature profiles and heat flux in the non-equilibrium steady state.
Towards that end, we couple the two ends of the system to a source and a drain of thermal energy represented by Langevin heat baths~\cite{langevin1908, pavliotis2014}.
Our results emphasize the importance of defects for heat transport properties in crystalline structures, and suggest to make use of the advantages of trapped ion systems to measure them.

The structure of the paper is as follows. Section~\ref{sec:ICC} 
presents the system under consideration. The corresponding dynamical equations are discussed in Sec.~\ref{sec:Dynamical}. The harmonic approximation valid at low energies is introduced in Sec.~\ref{sec:Linear}. In Sec.~\ref{sec:Temperature} we analyze how the presence of a topological defect affects the temperature distribution in the crystal, whereas the total heat flux is discussed in  Sec.~\ref{sec:HeatFlux}. Finally, we summarize our conclusions in Sec.~\ref{sec:Conclusions}.



\section{Ion Coulomb Crystals}
\label{sec:ICC}
We consider in the following $N$ ions of mass $m$ confined in a linear Paul trap that provides, in ponderomotive approximation, a harmonic potential with secular frequencies $\omega_z,\omega_x=\alpha\omega_z$ and $\omega_y = \beta\omega_x$. We neglect the micromotion from the fast oscillating electric field~\cite{major2005}. 
The system is characterized by the Hamiltonian:
\begin{align}
	H=\sum_i^N \frac{\vec{p}_i^{~2}}{2}+\mathcal{V}(\{\vec{r}_i\})\label{eq:H}
\end{align}
where $\vec{p}_i=(p_i^z,p_i^x,p_i^y)$ and $\vec{r}_i = (z_i,x_i,y_i)$ are, respectively, the momentum and position of the $i$-th ion.
The potential energy is of the form $\mathcal{V}=\sum_i v_i$, with
\begin{align}
2v_i(\{\vec{r}_j\})=z_i^2+\alpha^2\left(x_i^2+\beta^2y_i^2\right)+\sum_{j\neq i}\frac{1}{|\vec{r}_i-\vec{r}_j|}, \label{eq:potential}
\end{align}
where the last term is provided by the Coulomb repulsion between the ions. 
Throughout the paper we fix $\omega_z = 2\pi\times 25$kHz, and use $L=(e^2/4\pi\epsilon_0m\omega_z^2)^{1/3}$ as the length unit, $E=m\omega_z^2L^2$ as the energy unit,  
and $W=1/\omega_z$ as the time unit, with $e$ the elementary charge and $\epsilon_0$ the vacuum permittivity. 

The system crystallizes when the thermal energy of the ions is sufficiently low. 
The shape of the resulting crystal is determined by the competition between the Coulomb repulsion that tends to maximize the distance between ions, 
and the trap confinement, which pushes the particles closer together. 
Different structural phases have been observed depending on $N$ and the aspect ratios $\alpha$ and $\beta$. For the regime $\alpha,\beta>1$ considered in this work, 
the crystal structure lies solely in the $zx$ plane. 

Moreover, we tune $\alpha$ into a regime where the minimal-energy configuration is provided by the ions forming a triangular ladder 
along the $z$-axis, see Fig.~\ref{fig:scheme}. 
This crystal is commonly referred to as zigzag. 
Due to the mirror symmetry~($x_i\leftrightarrow -x_i$) of the potential $\mathcal{V}$, there exist two such states, which can be transformed into each other by flipping the positions along $x$.
Interestingly, this opens the possibility to introduce topological defects, or kinks, in the crystal, which can be interpreted as a domain wall between the two degenerate zigzag configurations, see Fig.~\ref{fig:scheme}.


Kinks in ion crystals have been subject to intensive study \cite{delcampo2010,partner2013,ulm2013}. It has been experimentally shown that a kink enables the emulation of nanofriction models, 
including a sliding to pinned transition, also known as Aubry transition, which
occurs due to the local incommensurability of the ion distances in the upper ($x>0$) and lower ($x<0$) sub-chains of the triangular ladder \cite{kiethe2017,kiethe2018}. 
The transition occurs when, by increasing $\alpha$, the crystal is squeezed closer to the $z$-axis, modifying the influence of the sub-chains on each other.  
Most importantly, when transitioning from the sliding to the pinned phase the $\mathds{Z}_2$ symmetry of the crystal along $z$ is broken, leading to robust localization features in the dynamics~\cite{timm2020}.
The defect slides into one of two possible equilibrium positions away from the trap center or, if the thermal energy permits it, jumps perpetually between the two configurations by overcoming the energy barrier that connects them. 
Although the dynamics is non-linear, the blockade of the energy transport can be traced back to the presence of asymmetric motional modes of the crystal that dominate the dynamics for small-enough energies. 
In the following, we study how the transition influences the thermal conductivity of the system when the crystal is transporting heat from a warmer to a colder bath.



\section{Dynamical equations}
\label{sec:Dynamical}
In order to investigate the thermal conductivity properties of a two-dimensional Coulomb crystal, we assume that the particles at the edges of the system are coupled to Langevin heat baths with different temperatures, as schematically indicated in Fig.~\ref{fig:scheme}. Therefore, the Hamilton equations determined from the Hamiltonian \eqref{eq:H} must be modified to include the corresponding dissipation and fluctuation terms.
The resulting Langevin equations acquire the form:
\begin{align}
\frac{d^2}{dt^2}\vec{r}_i=-\vec{\nabla}_i \mathcal{V}-\boldsymbol{\Gamma_i}\cdot\vec{p}_i+\vec{\xi}_i(t)\label{eq:dyn}
\end{align}
%
%
where $\boldsymbol{\Gamma_i} = \text{diag}(\gamma_i^z,\gamma_i^x,\gamma_i^y)$ is a diagonal matrix containing the dissipation rates in the different spatial dimensions. The stochastic force
$\vec{\xi}_i(t)$, provided by momentum kicks exerted by the heat baths, fulfills the fluctuation-dissipation theorem:
\begin{align}
\braket{\vec{\xi}_i(t)}=\vec 0\quad\braket{\vec{\xi}_i(t)\otimes\vec{\xi}_j(t')}=2\boldsymbol{\Gamma_i}\cdot \boldsymbol{T_i} \delta_{ij}\delta(t-t') \label{eq:fluc}
\end{align}
where $\braket{}$ denotes the ensemble average, ($\vec a\otimes\vec b)_{ij}=a_ib_j$ is the outer product of two vectors and 
$\boldsymbol{T_i}=\text{diag}(T_i^z,T_i^x,T_i^y)$ is a matrix containing the temperatures of the heat baths in units of $F=E/k_B$, with $k_B$ the Boltzmann constant.
In a trapped-ion experiment, the emulation of different heat baths may be accomplished by Doppler cooling lasers
detuned from a cooling transition~\cite{metcalf1999}.
While in this case the reachable temperatures of the heat baths are Doppler-limited, advanced cooling techniques are able to reach sub-Doppler regimes. 
We assume that the projection of the cooling lasers is the same for all spatial dimensions, hence we can write $\boldsymbol{\Gamma_i} = \gamma_i\mathds{1}$ and $\boldsymbol{T_i} = T_i\mathds{1}$. 

We are interested in the behavior of the dynamical temperature of the ions, which we define as
\begin{align}
\tau_i = \frac{\braket{\vec{p}_i^{~2}}}{3}, \label{eq:temperaturedef}
\end{align}
as well as in the total amount of energy the crystal can transport. 
To quantify the latter, we define the net energy the system gains from the coupling to the heat baths
\begin{align}
\frac{dH}{dt}=\sum_i j_i
\end{align}
where $j_i$ is the energy transported from the heat bath to ion $i$. 
Calculating the time-derivative of $H$ and inserting the Langevin equation~\eqref{eq:dyn}, we can write
\begin{align}
j_{i}=\vec{p}_i(t)\cdot\vec{\xi}_i-\vec{p}_i\cdot\boldsymbol{\Gamma_i}\cdot\vec{p}_i.
\end{align}

During the thermalization process, the crystal, dependening on the initial conditions, gains or looses energy. 
When the equilibrium steady state is reached, the same amount of energy is dissipated into the colder heat bath as is flowing into the system from the hotter bath, so that $\sum_i\braket{j_i}=0$.
The amount of energy that is transported, i.e. flowing into the system on one end and dissipated at the other end of the system, is the systems heat flux, defined by
\begin{align}
J(t)=\frac{1}{2}\sum_i|\braket{j_i(t)}| \label{eq:heatflux}
\end{align}
which we employ as a measure for the thermal conductivity of the crystal.
In the steady state, $lim_{t\rightarrow\infty}J(t)$ gives the amount of energy transported through the system. 
In order to calculate $J$, we employ Novikov's theorem~\cite{novikov1965}, which yields
\begin{align}
\braket{j_i}&=\text{tr}(\boldsymbol{\Gamma_i}\cdot \boldsymbol{T_{i}})-\braket{\vec{p}_i \cdot \boldsymbol{\Gamma_i} \cdot \vec {p}_i}\label{eq:flux}\\
&=3\gamma_i(T_i-\tau_i)
\end{align}
where the last equality is only valid for our choice $\boldsymbol{\Gamma_i},\boldsymbol{T_i}\propto\mathds{1}$.
While the first term in Eq.~\eqref{eq:flux} is externally determined, the second one characterizes the response of the system to the heat current and needs to be calculated. 
Towards this end, we perform numerical calculations solving the stochastic dynamical equations~\eqref{eq:dyn} for discretized time steps~\cite{kloeden1992}.
For a given set of parameters, we calculate $500$ independent trajectories of the ions.
In order to determine the ensemble averages in Eq.~\eqref{eq:temperaturedef}, we average over the trajectories and build a time average of $\SI{50}{\milli\second}$ when the system has reached the steady state. 
In addition to this numerical approach,  
the linearization of the dynamical equations 
provides important insights, as detailed below.



\section{Linear analysis}
\label{sec:Linear}
Assuming that the energy of the ions only allows for small fluctuations of their positions, we may expand the 
potential \eqref{eq:potential} up to second order in the deviations from their average positions.
This approximation permits on one side analytically-solvable dynamical equations, since the system is described by coupled harmonic oscillators. 
On the other side, when compared to full numerical computations, it reveals the relevance of the non-linearity induced by the Coulomb interaction in the heat transport in the crystal.

For vanishing temperature, the ions settle down at their equilibrium configuration $\vec{u}_0$, where we have condensed all degrees of freedom in a single state vector $\vec{u}=(\vec{r}_1,\dots,\vec{r}_N,\vec{p_1},\dots,\vec{p}_N)$.  We expand the dynamical equations \eqref{eq:dyn} up to first order in the deviations from the equilibrium $\vec{q}=\vec{u}-\vec{u}_0$, which yields
\begin{align}
\frac{d}{dt}\vec{q}=-\begin{pmatrix} 0 & -\mathds{1} \\ \boldsymbol{K} & \boldsymbol{\Gamma}\end{pmatrix}\cdot \vec{q}+\begin{pmatrix}0\\\vec{\xi}(t)\end{pmatrix}
\end{align}
where we have written in a compact way the dissipation-rate matrices $\boldsymbol{\Gamma}=\text{diag}(\boldsymbol{\Gamma_i})$ and the stochastic forces $\vec{\xi}(t)=(\vec{\xi}_1(t),\dots,\vec{\xi}_N(t))$.
The coherent dynamics is provided by the dynamical matrix $\boldsymbol{K}=\vec{\nabla}\otimes\vec{\nabla} ~\mathcal{V}(\{\vec r_i\})\big|_{\vec{u}_0}$. 
We diagonalize the dynamical matrix, $\boldsymbol{U}^T\cdot \boldsymbol{K} \cdot \boldsymbol{U}=\boldsymbol{D}$, where the diagonal matrix $\boldsymbol{D}$ and the unitary matrix $\boldsymbol{U}$, contain the eigenfrequencies of the crystal and the spatial structure of the corresponding eigenmodes. 
Denoting by $\vec{\theta}$ the state vector of the eigenmodes containing their amplitudes and momenta, we obtain an equivalent, more convenient, formulation of the Langevin equations:
\begin{align}
\frac{d}{dt}\vec{\theta}=\begin{pmatrix} \boldsymbol{U}^T &
 0 \\ 0 & \boldsymbol{U}^T\end{pmatrix}\cdot \frac{d\vec q}{dt} &= - \underbrace{\begin{pmatrix}
0 & -\mathds{1} \\
\boldsymbol{D} & \boldsymbol{\tilde\Gamma}
\end{pmatrix}}_{\boldsymbol{\Omega}}\cdot\vec\theta + \begin{pmatrix}
\vec 0 \\
\vec \Xi
\end{pmatrix} \label{eq:lindyn}
\end{align}
where $\boldsymbol{\tilde\Gamma}=\boldsymbol{U}^T\cdot\boldsymbol{\Gamma}\cdot \boldsymbol{U}$ is the transformed dissipation matrix, and $\vec\Xi=\boldsymbol{U}^T\cdot \vec\xi$ is the transformed stochastic force vector.
The latter fulfills a fluctuation-dissipation theorem as that of Eq.~\eqref{eq:fluc}, but with the transformed temperatures $\boldsymbol{\tilde T} = \boldsymbol{U}^T \cdot \boldsymbol{T} \cdot \boldsymbol{U}$. 
Note that the modified dissipation matrix and the transformed temperature matrix are not necessarily diagonal anymore, which can be interpreted as a dynamical coupling of the motional modes.
The Langevin equations \eqref{eq:lindyn} are formally solved by
\begin{align}
\vec\theta(t) &= e^{-\boldsymbol{\Omega} t} \cdot \vec\theta (0) +\int_0^te^{\boldsymbol{\Omega}(s-t)}\cdot\begin{pmatrix}
\vec 0 \\
\vec \Xi(s) \end{pmatrix} ds
\end{align}
where the first term describes the damped oscillations of the initial mode populations, whereas the latter part describes the stochastic motion. 
We insert this solution into the second moment matrix $\boldsymbol{C}(t)=\braket{\vec\theta(t)\otimes\vec\theta(t)}$, obtaining
\begin{align}
\boldsymbol{C}(t)&=e^{-\boldsymbol{\Omega} t}\cdot \boldsymbol{C}(0)\cdot e^{-\boldsymbol{\Omega}^T t} \nonumber \\ 
&+\int_0^t e^{\boldsymbol{\Omega} (s-t)}\cdot \begin{pmatrix} 0 & 0 \\ 0 & 2\boldsymbol{\tilde\Gamma}\cdot \boldsymbol{\tilde T} \end{pmatrix}\cdot e^{\boldsymbol{\Omega}^T (s-t)} ds,
\end{align}
from which we can read off the dynamical temperatures of the different motional modes $\tilde\tau_i =\boldsymbol{C}_{i+3N,i+3N}$, after carrying out the time integral.
Finally, we can calculate the net heat flux for each motional mode, which is given by 
\begin{align}
\braket{\tilde j_i} = \left(\boldsymbol{\tilde\Gamma}\cdot\boldsymbol{\tilde T}\right)_{i,i}-\sum_l^{3N}\boldsymbol{\tilde\Gamma}_{i,l}\boldsymbol{C}_{l+3N,i+3N}.\label{eq:modej}
\end{align}
The motional mode vectors are generally spatially extended and therefore the modes couple to the hotter and the colder bath simultaneously. 
We can think of the modes as harmonic oscillators coupled to two different thermal baths at the same time, and hence $\braket{\tilde j_i}$ vanishes in the steady state.
However, we can unambiguously split the dissipation matrix $\boldsymbol{\tilde\Gamma} = \boldsymbol{\tilde\Gamma}^h+\boldsymbol{\tilde\Gamma}^c$ and the temperature matrix $\boldsymbol{\tilde T} = \boldsymbol{\tilde T}^h+\boldsymbol{\tilde T}^c$ into the contributions coming from the different heat baths. 
This allows for the calculation of the heat flux from the hot bath into the motional modes, as well as of the heat flux dissipated into the colder bath, 
by inserting the respective parts of the matrices into Eq.~\eqref{eq:modej}.
In the steady state these two terms add up to zero so that the absolute value of one of them gives the energy transported by the mode $i$, the total flux \eqref{eq:heatflux} is then given by the sum over all modes.



\section{Temperature distribution}
\label{sec:Temperature}

\begin{figure}
\includegraphics[width=0.45\textwidth]{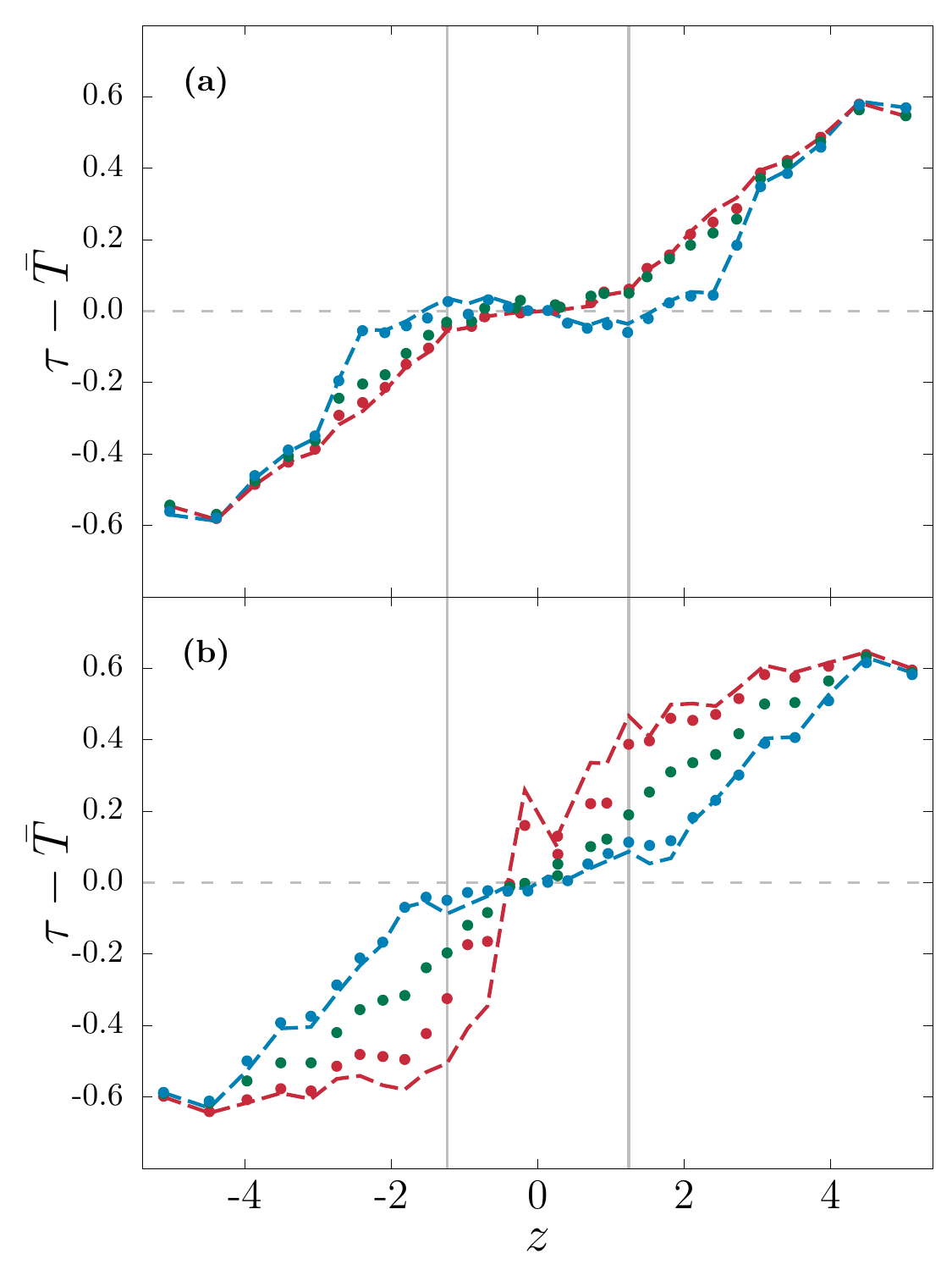}
\caption{Temperature profiles of an ion crystal of $N=30$ ions for the symmetric sliding phase $\alpha=6.0$~(a) and the symmetry-broken pinned phase $\alpha=7.0$~(b),
the graphs are normalized by $0.5(T^h-T^c)$. Results in blue are for the defect-less zigzag, whereas results for a crystal with a topological defect are shown in red and green. The dashed lines depict the results of the harmonic approximation, the circles indicate numerical results for the same parameters. For the blue and green curves the bath temperatures have been set to $\SI{0.5}{\milli\kelvin}$ and $\SI{0.7}{\milli\kelvin}$, whereas the graph in red is for $\SI{0.05}{\milli\kelvin}$ and $\SI{0.25}{\milli\kelvin}$. The vertical lines indicate the position of the ions that are used for the calculation of $dT$ (see text). }
\label{fig:temps}
\end{figure}

In this section, we investigate the temperature distribution in two-dimensional ion crystals in the presence of a topological defect, focusing on the impact of the symmetry breaking at the Aubry transition~\cite{kiethe2017}. 
Throughout this section, we consider that the four left-most and four right-most ions of the crystal are coupled to thermal baths, see Fig.~\ref{fig:scheme}, with fixed dissipation rate $\gamma/W=\SI{20}{\kilo\hertz}$ that is comparable to experimentally reached values~\cite{pyka2013}.
We consider a temperature difference between the two heat baths $(T^h-T^c)F=\SI{0.2}{\milli\kelvin}$, and set the average temperature $\bar T = (T^h+T^c)/2$ to different values in order to assess the effects of 
thermal fluctuations on the Aubry transition.
Figure~\ref{fig:temps} shows the  steady-state temperature distributions for different trap aspect ratios $\alpha$, for a zigzag crystal with and without a kink.



For a defect-free zigzag crystal with $\alpha=6.0$ we observe a sharp temperature edge at both ends of the crystal, and a flat profile for the inner ions, 
similar to the results observed in a linear ion chain~\cite{lin2011}. Changing $\alpha$ does not substantially change the profile, although for $\alpha=7.0$ the central ions show a slight temperature gradient.
The excellent agreement between the numerical results and the harmonic approximation indicates the irrelevance of non-linearities for these temperatures.
These findings differ from the results of Ref.~\cite{ruiz2014} for the same particle number which showed the emergence of a temperature gradient in the zigzag phase. 
We assign this discrepancy to the much larger temperatures of several $\si{\milli\kelvin}$ and stronger dissipation rates considered in that work. 
As pointed out in Ref.~\cite{ruiz2014}, thermal fluctuations lead to non-linearities being probed during the dynamics, which give rise to coupling and scattering of the phonon modes of the crystal.
Ultimately, the break-down of the harmonic description was pinpointed as the reason for the growth of the temperature slope by a Fourier analysis of the ion positions. 
Our results of the defect-free zigzag
complement this discussion since they show that 
the absence of a temperature gradient can be recovered for temperatures of the order of the Doppler temperature, and far away from the linear-zigzag transition.


The presence of a kink in the sliding phase smoothens the temperature profile, reducing the drops at the outer parts of the system, see Fig.~\ref{fig:temps}(a). This results from the localization of the spatial shape of the motional modes induced by the defect, that breaks the local translation invariance in the zigzag region.
%

The so-called Peierls-Nabarro (PN) potential enables a deeper insight into the properties of the kink~\cite{nabarro1947,partner2013}.
When understood as a quasiparticle inside the crystal, the defect moves inside an effective potential landscape that crucially depends on the trap configuration. 
In the sliding regime the defect is repelled from the edges of the inhomogeneous crystal, resulting in an approximately harmonic PN potential, with its minimum at the trap center. 
The deviations of the PN potential from its harmonic approximation are not probed at the considered temperature scale, and hence the presence of the defect does not result in significant nonlinear effects. As a consequence the steady state is well described by the linear theory, as seen in Fig.~\ref{fig:temps}(a).

The temperature profiles show a markedly different behavior when $\alpha$ is tuned into the pinning regime, as seen in Fig.~\ref{fig:temps}(b). Linear theory predicts a sharp drop of the temperature across the defect, and that the temperature profile does not present mirror symmetry, $\tau(z)-\bar T=\bar T - \tau(-z)$, as in Fig.~\ref{fig:temps}(a).
These observations are explained by the emergence of asymmetric modes in the spectrum, which are localized on one side of the defect, and hence are unable to contribute to the transport of heat across the system.
Since these modes are strongly coupled to only one thermal bath, their presence leads to a step-shaped temperature profile.

For low average temperatures, the linear-analysis prediction is supported by our numerical simulations. 
We observe a non-uniform temperature gradient across the crystal with the largest slope at the position of the defect. Although the qualitative observations agree, the markedly stronger deviations from the harmonic approximation compared to the sliding regime indicate the relevance of the nonlinear dynamics of the kink in the pinned phase, even well below Doppler temperature. When the energy scale of the baths is increased, the profile becomes close to a linear gradient such that no sharp feature of the energy blockade can be observed. 
This observation stands in clear contrast to the predictions of linear theory, marking the onset of nonlinear dynamics.

\begin{figure}
    \centering
    \includegraphics[width=0.48\textwidth]{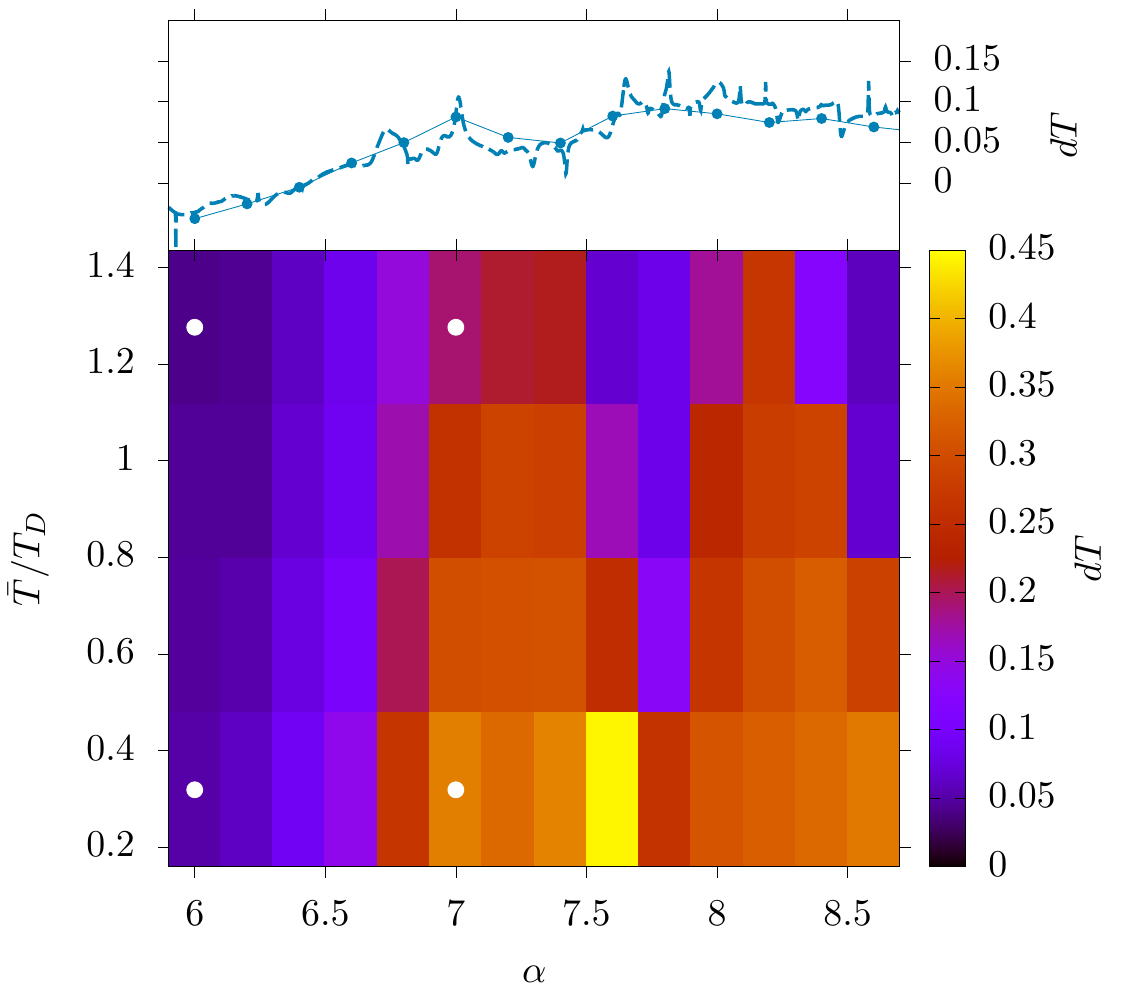}
    \caption{(top) Temperature difference between the $11$-th and the $20$-th ion normalized by the difference of the bath temperatures for a zigzag crystal. The dashed line depicts the linear theory result, and the circles indicate the numerical results for bath temperatures of $\SI{0.5}{\milli\kelvin}$ and $\SI{0.7}{\milli\kelvin}$. (bottom) Same quantity for a crystal initialized with kink as a function of the trap aspect ratio $\alpha$ and the mean temperature $\bar T$ of the baths, normalized by the Doppler temperature $T_D=\SI{0.5}{\milli\kelvin}$ for $\text{Yb}^+$. The white dots indicate the parameter choices from figure \ref{fig:temps}. For $\alpha>7.5$ the kink can be lost for large $\bar T$ (see text). }
    \label{fig:heatmap}
\end{figure}

As shown above, the steady-state temperature distribution provides a clear signal of the Aubry transition. To gain a better understanding of the effect of the Aubry transition and its interplay with thermal fluctuations, we depict in Fig.~\ref{fig:heatmap} the temperature difference $dT$ between the $11$-th and the $20$-th ion, as an indicator for the profile structure in the central region, see the vertical lines in Fig. \ref{fig:temps}. 
It is shown as a function of the trap aspect ratio $\alpha$ and the average temperature $\bar T$ of the two heat baths, blue regions indicate a close to vanishing temperature gradient whereas red to yellow marks strong temperature drops.
As a benchmark, we show in the upper diagram $dT$ as a function of $\alpha$ for a defect-free zigzag crystal. 
As discussed above, the zigzag crystal exhibits only a small gradient in the centre, not larger than $15\%$ of the temperature difference of the baths.

The lower plot of Fig.~\ref{fig:heatmap} depicts the results for a crystal initialized with a defect. In the sliding phase, $\alpha<6.4$, $dT$ is on the same order as for the defect-free zigzag crystal. This result is independent of the average temperature and agrees with the profiles shown in Fig.~\ref{fig:temps}.
Increasing $\alpha$ into the pinned phase, the temperature slope rises significantly at the critical point for small temperatures, clearly pinpointing the Aubry transition as the cause of the modification in the steady state distribution. Up to a value of $\alpha\approx 7.5$ there exists a parameter window in which the temperature slope for the central 10 ions covers around $40\%$ of the difference between the bath temperatures (note that the ions coupled to the thermal baths do not reach the bath temperatures due to the insufficient coupling strength $\gamma$, see Fig. \ref{fig:temps}). 



When the average temperature of the system is increased for a fixed $\alpha$ the system shows a transition into a phase with a weaker temperature slope for the defect ions, as observed in Fig.~\ref{fig:temps}(b).
The PN potential is again key to understand this feature. 
At the Aubry transition the globally confining, smooth PN potential develops periodic barriers~\cite{partner2013}.
Crucially for the discussion, a local maximum rises at the crystal center, leading to the emergence of two degenerate local minima located off the center.
For small energies, the defect falls spontaneously into one of these local minima, thereby breaking the symmetry of the crystal, and remains close to the randomly chosen configuration.
Hence, the dynamical properties are to a certain extent characterized by the harmonic approximation of the potential around that configuration, and phenomena like the sharp temperature drop emerge.

However, for large enough temperatures the defect exhibits a finite probability to overcome the energy barrier that emerged at the Aubry transition via thermal fluctuations, and can hence change its configuration between the two degenerate minimal energy states. Such a hopping results in the smoothening of the temperature profile~(see Fig.~\ref{fig:temps}(b)), and therefore in the reduction of $dT$, as it induces heat transport between spatial regions which remained disconnected at low energies. In this thermally-delocalized regime, the steady state $dT$ remains however larger than in the defect-free zigzag case, as a remaining signal of the symmetry breaking due to the finite dwelling time in one of the symmetry broken states. The observed thermal crossover resembles that induced by thermal fluctuations at the linear-to-zigzag structural transition~\cite{kiethe2021,li2019,delfau2013}, or by quantum 
fluctuations at the Aubry transition for much colder systems~\cite{timm2021,bonetti2021,borgonovi1989}. 

The critical temperature of the crossover is a non-trivial function of the trap aspect ratio.
While it increases with increasing $\alpha$ in the 
pinned phase, it saturates around $\alpha=7$ and later decreases for $\alpha> 7.5$. 
For larger values of $\alpha$ the kink undergoes a second transition into a localized shape, denoted as odd kink~\cite{partner2013,landa2013}. Most importantly, the PN potential changes into an inverted harmonic oscillator, while the periodic modulations introduced at the Aubry transition remain and stabilize the defect for small energies in local minima.
Since the crystal edges are now attracting the kink, a transition into a thermally-delocalized phase opens a channel for the complete loss of the kink. 
Indeed, the probability to travel to the crystal edges and vanish there becomes non-zero
when the kink is able to overcome the barriers between local potential minima in the PN potential and statistically hop between them. We observe the losses of the defect in our simulations for $\alpha>7.5$ but do not post-select those trajectories in which the kink stays confined for the calculation period. The reduction of the temperature gradient for large $\alpha$ and $\bar T$ observed in Fig.~\ref{fig:heatmap} is caused by the loss of the kink 
characterizing the delocalized phase in this regime.
The critical temperature for the thermal crossover to the delocalized regime drops to a local minimum at $\alpha=7.8$, coinciding with the crossover to the odd kink. 
Subsequently, the reduced heat transport becomes more robust again around $\alpha\approx 8.0$. The two regions with robust $dT$ observed in Fig.~\ref{fig:heatmap} match well the observed parameter windows of a strong blockade of a coherent excitation~\cite{timm2020}. For $\alpha>9.0$ the crystal with a kink is not a stable equilibrium of the system anymore as the modulations in the PN potential, which are crucial for the stability of the defect, decline in size for growing $\alpha$ and vanish subsequently.

One could expect that the crossover into the thermally-delocalized phase should occur when the temperature of the system becomes comparable to the PN barrier. Interestingly, although the energy barriers between different equilibrium states of the kink are typically of several mK~\cite{partner2013}, thermal delocalization occurs on the Doppler temperature scale~($T_D\approx \SI{0.5}{\milli\kelvin}$ here), as seen in Fig.~\ref{fig:heatmap}. 
Although we argue that the observed features can be understood from the form of the PN potential, 
this discrepancy indicates a more involved interplay, yet to be explored, between the thermal fluctuations of the kink and the residual motional modes of the crystal. 

\section{Heat flux}
\label{sec:HeatFlux}

\begin{figure}
\includegraphics[width=0.48\textwidth]{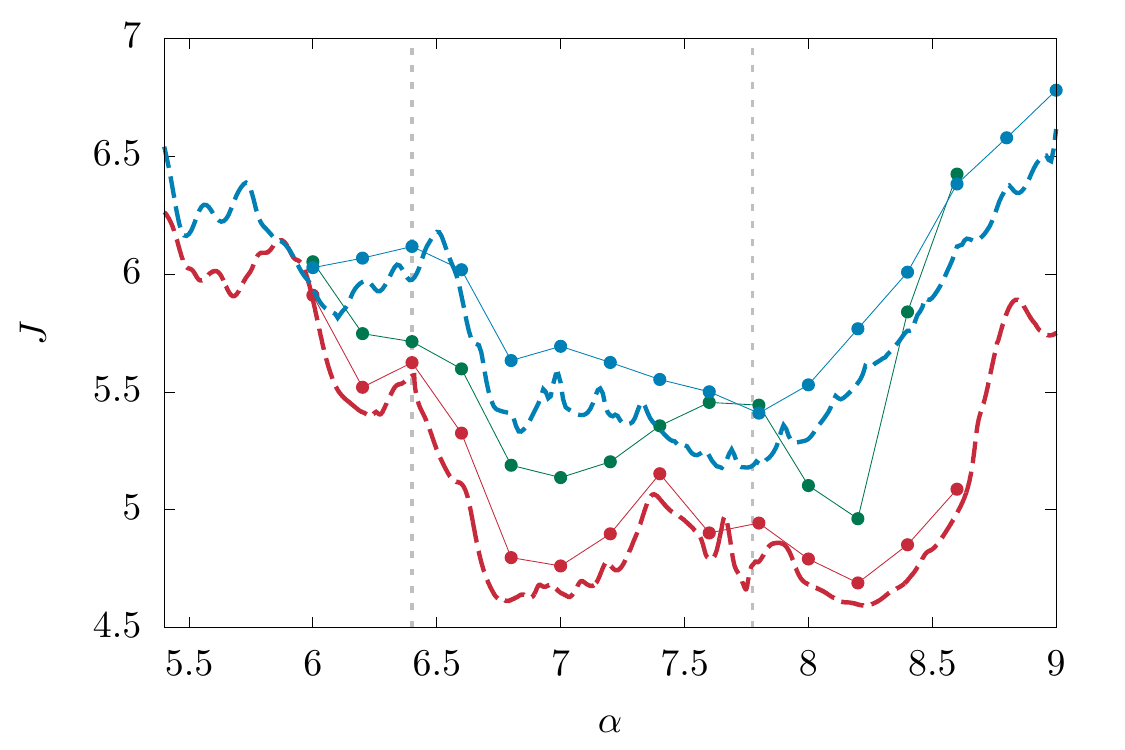}
\caption{Total heat flux $J$ through the crystal~(normalized by $0.5\gamma (T^h-T^c)$) as a function of the trap aspect ratio $\alpha$. The blue curves indicate the results for the defect-free zigzag crystal, while red and green graphs depict the case with a kink. The dashed lines indicate the linear theory results, and the circles our numerical results. For the latter the bath temperatures have been set to $\SI{0.5}{\milli\kelvin}$ and $\SI{0.7}{\milli\kelvin}$ for the blue and green graph, whereas in red we depict our results for $\SI{0.05}{\milli\kelvin}$ and $\SI{0.25}{\milli\kelvin}$. The gray dashed bars indicate the Aubry transition and the crossover to the odd kink. }
\label{fig:flux}
\end{figure}

We analyze at this point the total heat flux $J$, given by Eq.~ \eqref{eq:heatflux}, transported in the steady state, see Fig.~\ref{fig:flux}. For a defect-free zigzag crystal, both linear theory and the numerical results show a global minimum between $\alpha =6.5$ and $\alpha=8.5$. For larger $\alpha$, $J$ shows an approximately linear growth with $\alpha$, whereas 
for lower $\alpha$ it presents a more irregular growth.
Nonlinearities lead to a speedup of heat transport, since our numerical results 
show an uniform offset to larger values compared to linear theory. A similar dependence on $\alpha$ has been reported in Ref.~\cite{ruiz2014}, employing numerical simulations, although the presence of a trap configuration with minimal heat flux in the considered parameter window has not been discussed.
We address this point in the final part of this section.

For $\alpha>6.0$, the presence of the kink markedly reduces the amount of heat the crystal can transport, even in the sliding phase in which the crystals symmetry persists, see the red graphs in figure \ref{fig:flux}. The results for $J$ based on the harmonic approximation show in the sliding regime a heat flux slightly below the values for the defect-free case. At the Aubry transition, they display an abrupt decrease, followed by two minima of $J$, one at $\alpha\approx 7.0$ and the other at $\alpha\approx 8.25$.
These minima agree well with the regions of a robust temperature gradient observed in Fig.~$\ref{fig:heatmap}$.

For a low system temperature~(red circles in Fig.~\ref{fig:flux}), the numerical results remain close to the linear theory prediction, but, as for the defect-free case, they show an uniform offset 
towards faster transport. For larger temperatures, the kink becomes thermally delocalized and hence allows for faster energy transport~(green circles in Fig.~\ref{fig:flux}). Although the signal strength of the Aubry transition is therefore reduced, the decrease of $J$ remains, even for these temperatures, 
a clear signature of the transition.



As a final point, we address a subtle issue concerning the presence of a minimal heat flux for the zigzag crystal. 
In Ref.~\cite{ruiz2014}, the reduction of $J$ when $\alpha$ is quenched through the linear-to-zigzag transition is explained by the growing inter-ion distances~\cite{ruiz2014} when the ions start exploring the radial~($x$) direction. This argument cannot explain however our results, exhibiting a minimum in the heat flux for the defect-free zigzag crystal, as the ion distances are a monotonously-growing function with decreasing $\alpha$. Furthermore, the linear theory results show finer features of the $J$ curve, as shown in Fig.~\ref{fig:flux}.

\begin{figure}
    \centering
    \includegraphics[width=0.48\textwidth]{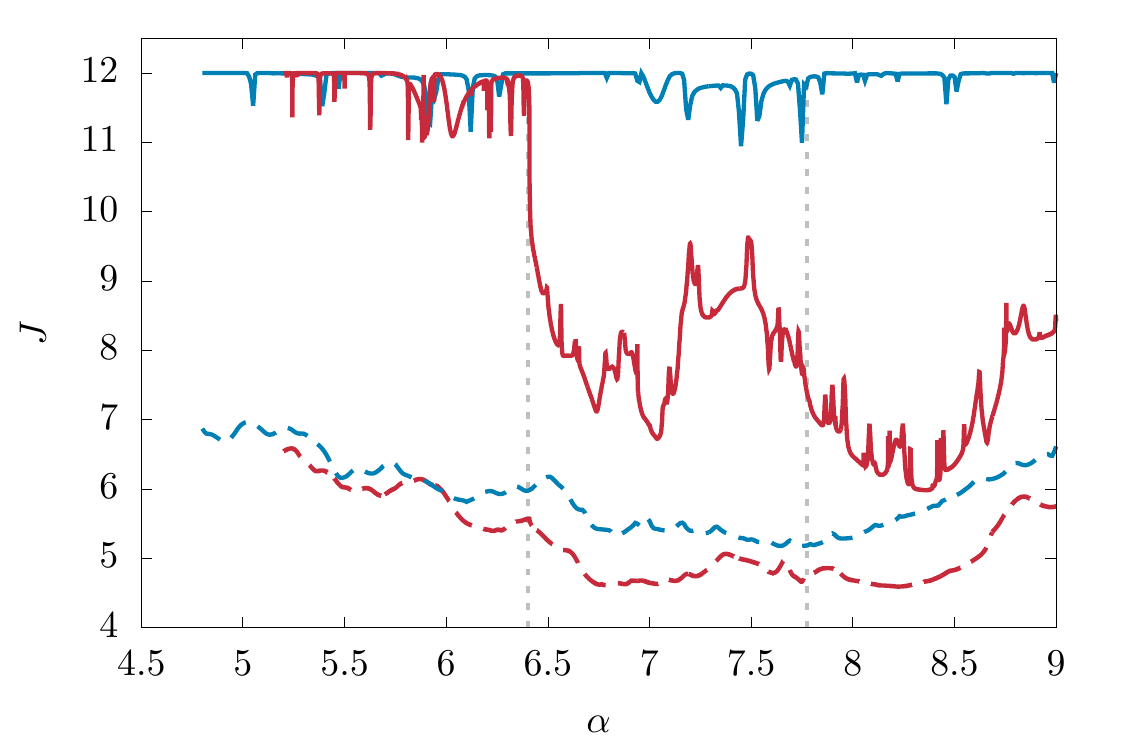}
    \caption{Normalized total heat flux obtained from linear theory for different values of the coupling $\gamma$. The dashed lines indicate the results without~(blue) and with~(red) kink for $\gamma/W=\SI{20}{\kilo\hertz}$, as in Fig.~\ref{fig:flux}, whereas the solid lines depict the zero coupling limit, calculated for $\gamma/W=\SI{0.002}{\hertz}$. The vertical dashed lines indicate again the points of Aubry transition and the crossover to the odd kink.}
    \label{fig:resonance}
\end{figure}

In order to argue that both these observations can be explained from the structure of the motional modes, we compare in Fig.~\ref{fig:resonance} the result for the heat flux observed in Fig.~\ref{fig:flux} with the outcome of the calculation in the limit of vanishing coupling $\gamma$. In this discussion, we only compare the linear theory results, since the numerical calculation for $\gamma\rightarrow 0$ demands unreachable equilibration times.
We believe this approximation is however enough for the following discussion, since our results exhibit a good agreement with the numerics for the case of a defect-free zigzag crystal. Note, that the results are normalized by $J_0=0.5\gamma (T^h-T^c)$ such that the trivial linear scaling of the flux with the coupling strength is accounted for. Changes in the heat conductivity are hence due to the  altered system response.

For the symmetric crystals, i.e. the zigzag and the kink in the sliding phase, 
the heat flux is independent of $\alpha$ and shows additional negative peaks.
The Aubry transition leads to a strong heat flux reduction in the presence of the kink,  but the sharp features persist in the pinned regime.
While the effect of the sliding-to-pinning transition remains observable 
for the value of $\gamma/W=\SI{20}{\kilo\hertz}$ employed above, the resonance features observed at low $\gamma$ were partially washed out. The fine modulations coincide with $\alpha$ values for which two motional modes contributing to the energy transport become degenerate, resulting in oscillations with a fixed phase relation. 
Since the mode vectors have alternating symmetry~(symmetric-antisymmetric under mirror transformation $z\leftrightarrow -z$) with increasing energy, for the symmetric crystal cases their correlated motion leads to a destructive interference in one half of the crystal. 
Ultimately, this results in a sharp decrease in heat conductivity at the resonance points.
For the crystal with broken symmetry, the mode vectors do not possess a fixed symmetry anymore and hence a resonance can increase energy transport, as the presence of positive peaks in $J$ in this phase only shows. 
Similarly as in a driven harmonic oscillator, the increase of the damping $\gamma$ broadens the resonance peaks, which subsequently start to overlap and finally wash out.

Based on these findings, we understand the reduction of the heat flux in zigzag crystals and the presence of a minimal conductivity configuration as a consequence of the density of mode crossings in the motional spectrum of the system in the considered temperature scale.
This is supported by the fact that the calculated heat flux is only weakly dependent on $\alpha$ when the crystal forms a chain~\cite{ruiz2014}.
In that phase the absence of mode frequency crossings as a function of $\alpha$ yields an invariant flux, and a small decrease with $\alpha$ can be explained due to the bunching of the radial mode frequencies. 
For off-resonant values of $\alpha$ each of the four ions coupled to one of the heat reservoirs contributes $J_0$ to the total heat flux $J$ for each dimension with $\gamma_i^\mu\neq 0$, which yields in our case $J=12 J_0$, a result that can be rigorously shown for harmonic oscillator models in the limit $\gamma\rightarrow 0$~\cite{rieder1967,lepri2003}.




\section{Conclusion}
\label{sec:Conclusions}

The presence of a topological soliton strongly affects heat transport in an ion crystal. A sloped temperature profile emerges when the defect is driven across the Aubry transition from the sliding to the pinning phase, whereas the defect-free system exhibits a negligible temperature gradient for temperatures close to the Doppler limit. Moreover, the nonlinear dynamics of the defect inside the crystal is exposed when the average temperature of the system increases, resulting in a thermally delocalized phase, and in the reduction of the effects of the Aubry transition. Our calculation of the heat flux clearly 
shows that the presence of the defect significantly hinders the energy transport, especially in the pinning phase, and demonstrate the importance of the motional mode spectrum due to the presence of resonances.

In addition, we encountered results that remain unanswered and require further investigation. The interplay between the motion of the defect inside the effective Peierls-Nabarro potential and the residual degrees of freedom of the crystal proves to be non-trivial, as the mismatch between the energy scales for the crossover to the delocalized regime and the size of the Peierls-Nabarro barriers localizing the kink demonstrates.
Here the formulation of the dynamical equations in a collective excitation formulation could reveal the influence of these two different types of degrees of freedom onto each other~\cite{willis1986,igarashi1989}.

Another aspect that has not been treated in this work is the dependence of the temperature profiles and heat flux on the system size. 
Especially the scaling of $J(N)$ is of interest since a deviation from a $\propto 1/N$ scaling would indicate a diverging thermal conductivity in a properly defined thermodynamic limit, keeping the ion density constant.
This analysis could also reveal the importance of finite size effects for the temperature gradients, possibly recovering the temperature steps at the contacts to the heat baths in the pinned phase.
Moreover, the response of the system to the temperature stress in a setup with a different axial confinement restoring (partly) the translational invariance could differ qualitatively from the results presented here. 
The inhomogeneous ion density in the crystals with harmonic confinement ultimately leads to the localization of the defect in the central region, which hinders its abilities to transport energy.

In addition, the expansion of the investigation of heat transport to other geometries, such as disc-shaped two-dimensional or cigar shaped three-dimensional systems could shed light onto the influence of dimensionality, making use of the versatility of trapped ion crystals~\cite{nigmatullin2016,podolsky2016,arnold2022}
Finally, we expect that the altered excitation dynamics in crystals with defects impact the sympathetic cooling of these systems as the observed reduced heat flux could lead to regions of weak dissipation rates.
The latter would be of direct importance for experiments investigating the properties of topological defects. \\

\acknowledgements
This project has been funded by the Deutsche Forschungsgemeinschaft (DFG) under Germany’s Excellence Strategy - EXC-2123 QuantumFrontiers–390837967 and through SFB 1227 (DQmat) – Project-ID 274200144, project A07.

\bibliographystyle{apsrev4-1}
\bibliography{HeatingCoolingPaper.bib}

\end{document}